\documentclass[superscriptaddress,twocolumn,prb,aps,10pt,longbibliography]{revtex4-1}

\usepackage{amsfonts,amssymb}
\usepackage[sumlimits,intlimits]{amsmath}
\usepackage{graphics}
\usepackage{graphicx}
\usepackage{mathrsfs}
\usepackage{textcomp}
\usepackage{verbatim}
\usepackage{hyperref}    
\usepackage{bm}          

\pdfoutput=1
	
\newcommand{\be}{\begin{equation}}
\newcommand{\ee}{\end{equation}}
\newcommand{\e}{\varepsilon}
\newcommand{\lb}{\ell_B}
\newcommand{\ab}{a_B}

\newcommand{\ibias}{I_\text{bias}}
\newcommand{\beql}{B_\text{EQL}}
\newcommand{\ignore}[1]{}

\begin{document}

\title{Spatially inhomogeneous electron state deep in the extreme quantum limit of strontium titanate}

\author{Anand Bhattacharya}
\thanks{These two authors contributed equally.}
\affiliation{Materials Science Division, Argonne National Laboratory, Argonne, Illinois  60439, USA}

\author{Brian Skinner}
\thanks{These two authors contributed equally.}
\affiliation{Materials Science Division, Argonne National Laboratory, Argonne, Illinois  60439, USA}
\affiliation{Massachusetts Institute of Technology, Cambridge, MA  02139, USA}

\author{Guru Khalsa}
\affiliation{Center for Nanoscale Science and Technology,
National Institute of Standards and Technology, Gaithersburg, Maryland 20899, USA}

\author{Alexey V. Suslov}
\affiliation{National High Magnetic Field Laboratory, Tallahassee, Florida  32310, USA}

\date{\today}

\begin{abstract}

When an electronic system is subjected to a sufficiently strong magnetic field that the cyclotron energy is much larger than the Fermi energy, the system enters the ``extreme quantum limit" (EQL) and becomes susceptible to a number of instabilities.  Bringing a three-dimensional electronic system deeply into the EQL can be difficult, however, since it requires a small Fermi energy, large magnetic field, and low disorder.  Here we present an experimental study of the EQL in lightly-doped single crystals of strontium titanate, which remain good bulk conductors down to very low temperatures and high magnetic fields.  Our experiments probe deeply into the regime where theory has long predicted electron-electron interactions to drive the system into a charge density wave or Wigner crystal state.  A number of interesting features arise in the transport in this regime, including a striking re-entrant nonlinearity in the current-voltage characteristics and a saturation of the quantum-limiting field at low carrier density.  We discuss these features in the context of possible correlated electron states, and present an alternative picture based on magnetic-field induced puddling of electrons.

\end{abstract}

\maketitle

When subjected to a sufficiently strong magnetic field, the bulk properties of an electronic system change dramatically.  In particular, a three-dimensional electron gas acquires very different behavior when the magnetic field $B$ becomes large enough that the cyclotron energy $\hbar \omega_c = \hbar e B/m$ exceeds the Fermi energy $E_F \propto \hbar^2 n^{2/3}/m$\ignore{, or, equivalently, when the magnetic length $\lb = \sqrt{\hbar/eB}$ becomes shorter than the Fermi wavelength $\propto n^{-1/3}$}.  (Here, $\omega_c$ is the cyclotron frequency, $\hbar$ is the reduced Planck constant, $-e$ is the electron charge, $m$ is the effective electron mass, and $n$ is the electron density.)  In this ``extreme quantum limit'' (EQL) electron motion in the directions perpendicular to the magnetic field is quantized, and all electrons occupy only the lowest Landau level. Semi-classically, one can visualize the EQL as the state in which electron trajectories are tight spirals along the field direction, with gyration radius $\lb$ that is much shorter than the typical inter-electron spacing.

\begin{figure}[htb]
\centering
\includegraphics[width=0.5 \textwidth]{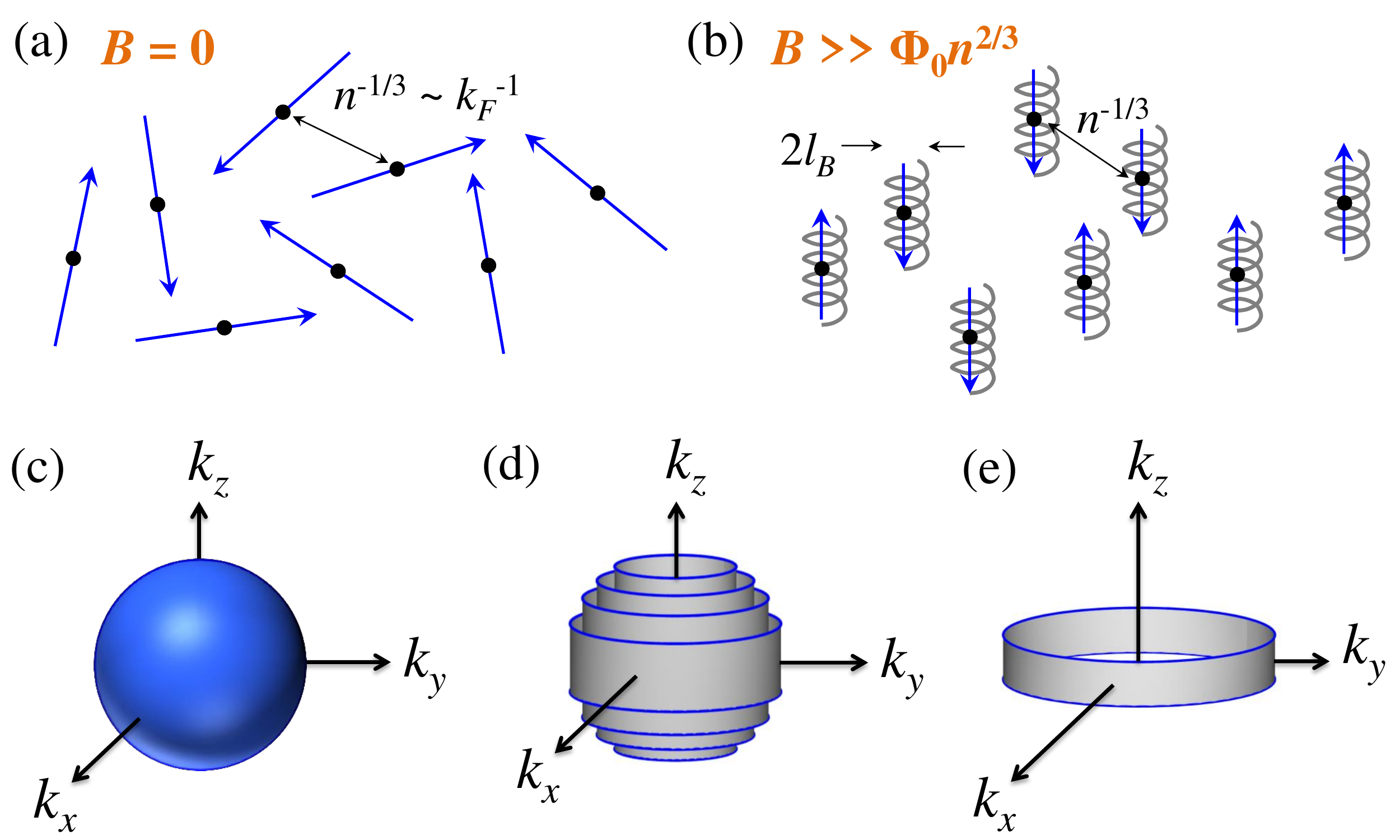}
\caption{ (a)--(b) Semi-classical picture of the trajectories of Fermi level electrons, both (a) at zero magnetic field, and (b) in the EQL.  $\Phi_0 = \pi \hbar/e$ is the flux quantum.  (c)--(e) show the evolution of the Fermi surface (blue areas) in a clean system with increasing magnetic field.  Gray areas show occupied electron states below the Fermi level.  (c) The usual spherical Fermi surface at $B = 0$.  (d) In a strong magnetic field, electron states are quantized into ``Landau cylinders", each with constant magnitude of the squared transverse momentum, $k_x^2 + k_y^2$.  (e) In the EQL, all electrons belong to the lowest Landau level.}
\label{fig:schematics}
\end{figure}

As a consequence of this quantization by magnetic field, in the EQL the electron kinetic energy retains a dependence only on the momentum parallel to the field, and the Fermi surface takes the form of two parallel rings in momentum space [see Fig.\ \ref{fig:schematics}(e)].  For a clean, low-temperature electron gas, such a dimensionally reduced dispersion implies a number of potentially competing instabilities, including spin or valley density wave, charge density wave (CDW), and Wigner crystallization.\cite{celli_ground_1965, kaplan_electron_1972, kleppmann_wigner_1975, halperin_possible_1987}

Such a high-field situation, however, is difficult to realize experimentally.  For a typical metal, for example, the EQL requires fields on the order of $10^5$\,T, and is thus relevant only for extreme astrophysical settings.\cite{lai_matter_2001, kaplan_electron_1972}  Realization of the EQL in the laboratory requires a material that can exhibit metallic behavior at very low electron density $n$.

Doped bulk strontium titanate, SrTiO$_3$ (STO), is such a material.  STO, a semiconducting perovskite oxide, has been studied intensively in recent years, with much attention devoted to its potential in thin films and heterostructure devices.\cite{haeni_room-temperature_2004, ohtomo_high-mobility_2004, thiel_tunable_2006, caviglia_electric_2008, son_epitaxial_2010}  But as a bulk material STO has attracted interest for over half a century,\cite{cowley_lattice_1964, barrett_dielectric_1952, frederikse_shubnikovhaas_1967, yamada_neutron_1969, neville_permittivity_1972, kahn_electronic_1964, \ignore{rice_persistent_2014,}schooley_superconductivity_1964} largely because of its anomalous dielectric response at low temperature.\cite{cowley_lattice_1964, yamada_neutron_1969, neville_permittivity_1972}  Indeed, STO has a static, long-wavelength dielectric constant $\e$ that reaches $\approx 24\,000$ at low temperatures (with a weak directional dependence\cite{neville_permittivity_1972})\ignore{, and arises from an incipient ferroelectric transition that is suppressed by quantum fluctuations}.  One implication of this enormous dielectric constant is that the effective Bohr radius $\ab = 4 \pi \e_0 \e \hbar^2/me^2$ associated with shallow donor states becomes extremely large: $\ab \approx 760$\,nm.  Consequently, in the absence of compensating acceptors, even a very small concentration of electron donors is sufficient to ensure that STO is on the conducting side of the Mott criterion, $n^{1/3}\ab \gtrsim 0.2$.\cite{mott_metal-insulator_1968}  Importantly, as we discuss below, this large dielectric constant also implies a significant robustness against localization by charge disorder.

In this paper we present a clear experimental realization of a deep-EQL state in STO.  Using transport measurements at low temperatures and high magnetic fields, we probe deeply into the EQL in a number of low-carrier-density samples.  
Our samples remain good bulk conductors throughout our measurement conditions, and at large magnetic fields they exhibit a strong, re-entrant nonlinearity in the resistivity. 
This nonlinearity is discussed in the context of possible correlated electron states, and an alternate picture is presented based on puddling of electrons in disorder potential wells.

Crucial to our study is the existence of a strong hierarchy of energy scales:
\be 
\hbar \omega_c \gg E_F \gg \mathcal{R}_E, \, k_BT.
\label{eq:hierarchy}
\ee 
Here, $\mathcal{R}_E = e^2/(8\pi \e_0 \e a_B)$ is the effective Rydberg energy and $k_BT$ is the thermal energy.  The first inequality in Eq.\ (\ref{eq:hierarchy}) is equivalent to the large magnetic field condition explained above, while the second set of inequalities guarantees that the electron state deep in the EQL is not destroyed either by thermal excitation or by freezeout of electrons onto donor impurities.  While previous high-field studies have managed to approach or even enter the EQL in doped semiconductors, simultaneously achieving both sets of inequalities in Eq.\ (\ref{eq:hierarchy}) has been more challenging.  For example, studies of the EQL in narrow band gap semiconductors, such as $n$-type InAs,\cite{Zeitler_magneto-quantum_1994} InSb,\cite{murzin_quasi-one-dimensional_2000, shayegan_magnetic-field-induced_1988, mani_possibility_1989, Zeitler_magneto-quantum_1994} and HgCdTe,\cite{shayegan_magnetic-field-induced_1988, mani_possibility_1989} are generally limited to the case where $E_F$ and $\mathcal{R}_E$ are similar in magnitude.  Consequently, in these materials electrons freeze onto donor impurities shortly after the EQL is reached, and achieving a deep-EQL electron state is not possible.  Previous high-field studies of STO have also generally failed to satisfy this hierarchy, either because the temperature was too high to satisfy the second inequality\cite{kozuka_vanishing_2008} or because the Fermi energy was too high to satisfy the first.\cite{allen_conduction-band_2013, lin_fermi_2013}


In our study, we examine millimeter-sized STO single crystals (dimensions: $\approx 7.5\text{\,mm} \times 7.5\text{\,mm} \times 0.5$\,mm), obtained from CrysTec GmbH.\cite{crystec}$^,$\footnote{This commercial product is described in this paper in order to specify adequately the experimental procedure. In no case does such identification imply recommendation or endorsement by the National Institute of Standards and Technology, nor does it imply that it is necessarily the best available for the purpose.}  
In order to introduce a finite concentration of conduction-band electrons, the samples were heated within a vacuum chamber following the protocol described in the Supplementary Information (SI).  This heating process is known to produce oxygen vacancies within the sample volume, which act as electron donors.\cite{\ignore{pasierb_comparison_1999, meyer_observation_2003,} de_souza_behavior_2012, paladino_oxidation_1965}$^,$\footnote{In principle, a single oxygen vacancy acts stoichiometrically as a double-donor, but it is generally accepted that one of the two electron states remains tightly bound to the doubly-charged oxygen ion, so that the vacancy donates only a single electron to the conduction band.\cite{\ignore{hou_defect_2010, lin_electron_2013, lopez-bezanilla_magnetism_2014,} janotti_vacancies_2014}}  The heating temperature was varied from one sample to another in order to produce samples with different doping levels.  For the samples that are the focus of this study, the resulting carrier densities ranged from $(7.7 \pm 1.1) \times 10^{15}$\,cm$^{-3}$ to $(1.5 \pm 0.2) \times 10^{18}$\,cm$^{-3}$.  (Here and below, all listed values of the experimental uncertainty correspond to $95\%$ confidence intervals.) 

Importantly, in addition to the oxygen vacancies, as-grown STO crystals are known to have a significant concentration of additional impurities, mostly Fe and Al, that act as deep acceptors.\cite{ensign_shared_1970, morin_energy_1973}  Indeed, previous studies based on chemical analysis and secondary ion mass spectroscopy\cite{de_souza_behavior_2012}  have shown these impurities to be present at the level of $\approx 7 \times 10^{17}$\,cm$^{-3}$.  Since this concentration is significantly larger than the measured carrier density $n$, our samples can be described as almost-completely-compensated semiconductors, for which the number of donors and acceptors are nearly identical and thus the total concentration of charged impurities $N_i$ is much larger than $n$.  This relatively large concentration of impurities is also consistent with the measured zero-field mobility, which suggests an impurity concentration on the order of $10^{18}$\,cm$^{-3}$ (as shown in SI Sec.\ II).  The consequences of the impurity concentration $N_i \gg n$ for transport are discussed in detail below.

\begin{figure}[htb]
\centering
\includegraphics[width=0.5 \textwidth]{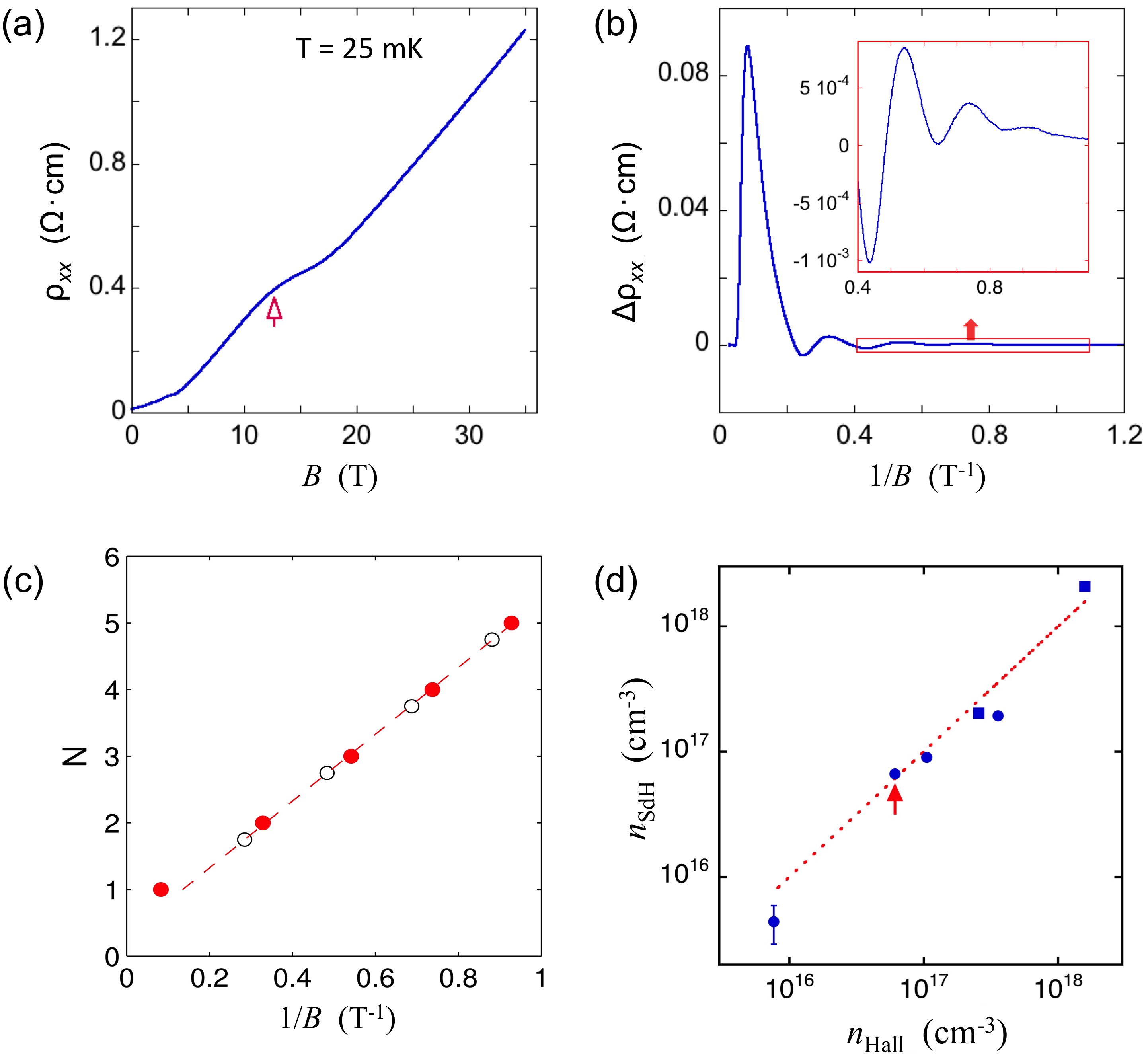}
\caption{Identifying the EQL via SdH oscillations.  (a) The longitudinal resistivity $\rho_{xx}$ at $T = 25$\ mK for one of our samples is plotted as a function of magnetic field.  The arrow indicates the onset of the EQL.  (b) The SdH oscillations of resistivity are more easily visible if one subtracts a smooth, fourth-order polynomial from the curve in (a).  (c) The index $N$ is plotted against $1/B_N$.  The value of $B_N$ can be identified either using the local resistance maxima (filled red circles) or the local minima of the derivative $d\rho/dB$ (open circles).  Experimental uncertainty in $B_N$ is $\approx 0.2$\,T, so that the uncertainty in $1/B_N$ is smaller than the symbol sizes.  For the open circles, the plotted value of $N$ is shifted by $1/4$.  (d) The carrier density $n_\text{SdH}$ inferred from the SdH period [see Eq.\ (\ref{eq:SdH})] is plotted against the measured Hall carrier density $n_\text{Hall}$.  Uncertainty in the value of $n_\text{Hall}$ is about $15\%$, and is reflected by the size of the symbols, while uncertainty in the value of $n_\text{SdH}$ is $\approx 5\%$ except where indicated.  Circles correspond to samples for which transport was measured in the $001$ direction, while squares indicate measurements in the $111$ direction. The arrow indicates the carrier density corresponding to the sample in parts (a)-(c).}
\label{fig:EQL}
\end{figure}

As the magnetic field is increased from zero, the longitudinal resistivity exhibits Shubnikov-de Haas (SdH) oscillations, as higher Landau levels are pushed outside the Fermi surface with increasing field.  In particular, the field $B_N$ at which the $N$th Landau level becomes depopulated follows\cite{abrikosov_alexei_a_fundamentals_1988}
\be 
\frac{1}{B_{N+1}} - \frac{1}{B_N} \equiv \Delta(1/B) \simeq \left( \frac{16}{9 \pi} \right)^{1/3} \frac{n^{-2/3}}{\Phi_0},
\label{eq:SdH}
\ee
where $\Phi_0 = \pi \hbar/e$ is the flux quantum, and the second equality in Eq.\ (\ref{eq:SdH}) corresponds to the usual limit of large $N$.  As shown in Fig.\ \ref{fig:EQL}, the observed oscillations of resistivity are periodic in $1/B$, suggesting a single, small electron pocket of Fermi surface.\cite{lin_scalable_2015}  The position of the final ($N = 1$) oscillation indicates that the EQL is reached at fields $B \approx 10$\,T to $20$\,T, varying from sample to sample.  Here the field $B_N$ is defined experimentally as the position in magnetic field of the $N$th local maximum of resistance, counted in order of decreasing magnetic field.  The identification of these maxima is facilitated by subtracting a smooth, fourth order polynomial from the $\rho$ vs.\ $B$ curve, as shown in Fig.\ \ref{fig:EQL}(b) (see also SI Sec.\ IV).  The period of oscillation is also confirmed by the periodicity of the first derivative of $\rho$ vs.\ $1/B$ [see Fig.\ \ref{fig:EQL}(c)].  Plotting $1/B_N$ against the Landau index $N$, as shown in Fig.\ \ref{fig:EQL}(c), indicates unambiguously that each of our samples is well within the EQL at our largest magnetic fields.  
\footnote{One can note that Eq.\ (\ref{eq:SdH}) assumes that Landau levels are not spin-degenerate.  The agreement of this equation with our measurements suggests that this is indeed the case in our samples for all appreciable magnetic fields.  This non-degeneracy can be expected if one assumes that the electron $g$-factor in STO is of order unity.  In this case the Zeeman energy is of order $100$\,$\mu$eV per Tesla of field, while the Fermi energy in our samples is in the range $100$\,$\mu$eV to $500$\,$\mu$eV, so that the conduction electrons become completely spin polarized even at relatively high Landau levels.}

\begin{figure}[htb]
\centering
\includegraphics[width=0.5 \textwidth]{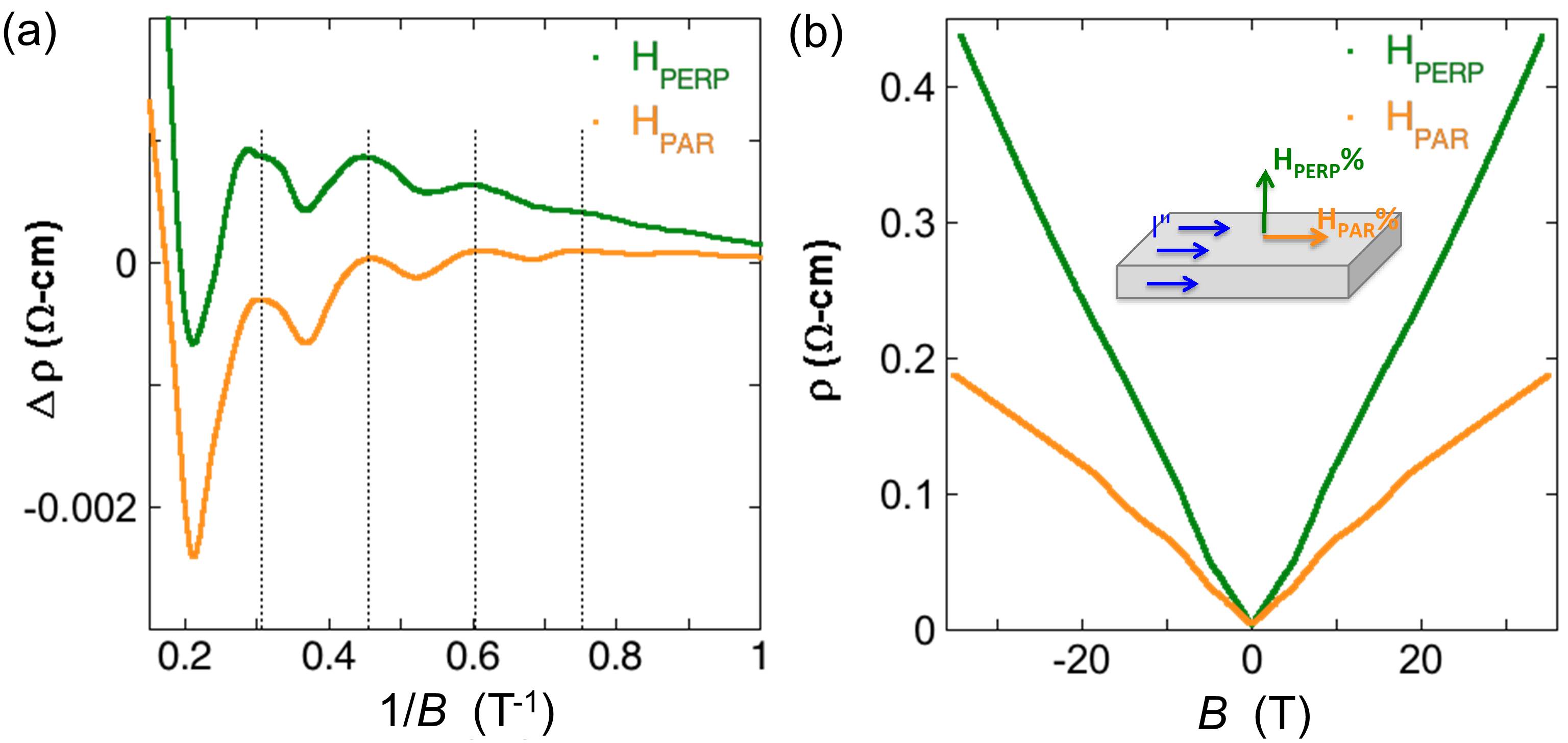}
\caption{Dependence of the resistivity on field direction.  (a) Shows that the period of the SdH oscillations is independent of whether the field is applied in the parallel or perpendicular direction.  (b) Plots the resistivity at $T = 25$\,mK as a function of magnetic field for both field  directions.  At large $B$ we observe linear MR and a significant anisotropy.}
\label{fig:n}
\end{figure}

Using Eq.\ (\ref{eq:SdH}), the value of the SdH period $\Delta(1/B)$ gives a measure of the electron density $n$.  The excellent linear fit to Eq.\ (\ref{eq:SdH}) indicates that the large-$N$ approximation is appropriate for all $N > 1$.  Further, the inferred value of $n$ closely matches the value obtained from Hall effect measurements, as shown in Fig.\ \ref{fig:EQL}(d).  The agreement between the two measures of $n$ suggests that electrons are uniformly distributed through the bulk of the STO crystal, since the SdH measurements are sensitive to the average Fermi energy of electrons in the sample, while the Hall effect observes only the thickness-averaged number of carriers.  The period of the SdH oscillations is also independent of whether the magnetic field is aligned parallel or perpendicular to the current, as shown in Fig.\ \ref{fig:n}(a), which confirms the three-dimensional nature of the electron system.  We do, however, observe a noticeable anisotropy in the magnetoresistance (MR) at large fields, as shown in Fig.\ \ref{fig:n}(b).

Our samples also exhibit a large linear MR in the EQL, as evidenced in Figs.\ \ref{fig:EQL}(a) and \ref{fig:n}(b), which show a MR ratio $\rho(B = 45\,\textrm{T})/\rho(B = 0) > 100$.  Such non-saturating, linear MR is commonly associated with semi-classical drift of electron orbits along contours of a disorder potential with a long correlation length,\cite{murzin_electron_2000\ignore{, murzin_quasi-one-dimensional_2000}, song_guiding_2015} or with strong spatial inhomogeneity of the carrier concentration or mobility.\cite{dykhne_anomalous_1971, parish_classical_2005, kozlova_linear_2012\ignore{, narayanan_linear_2015}}  Below we provide an additional comment on these possibilities.


While the Fermi energy of a three-dimensional electron gas is essentially constant at small values of the magnetic field, once the EQL is reached (after the last SdH oscillation) the Fermi energy $E_F$ acquires a strong dependence on the field strength.  In particular, in the EQL $E_F \propto \hbar^2 n^2 \lb^4/m \propto 1/B^2$, as constriction of electronic wave functions in the perpendicular directions reduces their quantum overlap and causes the Fermi energy to drop.  In our samples, $E_F$ is small enough that in the EQL only the lowest $t_{2g}$ band is relevant,\cite{khalsa_theory_2012, allen_conduction-band_2013} and this is consistent with the single frequency observed in the low field SdH measurement.\footnote{This lowest band has a slight anisotropy of the effective mass,\cite{allen_conduction-band_2013} resulting from the tetragonal distortion of the STO lattice at low temperature, so for the sake of making numerical estimates below we take the effective mass $m$ to be equal to the geometric mean of the three perpendicular masses, which gives $m \approx 1.7 m_0$, where $m_0$ is the bare electron mass.}

As the Fermi energy decreases with increasing field, the relative strength of the Coulomb interaction grows, as described by the ratio $\alpha = E_C/E_F$, where $E_C \propto e^2 k_F/4\pi\varepsilon_0\e$ is the typical strength of the Coulomb interaction between neighboring electrons in the field direction, and $E_F \propto \hbar^2 k_F^2/m$ is the Fermi energy.  Here, $k_F \propto n \lb^2$ is the Fermi wave vector in the field direction, so that one can write $\alpha = (n \lb^2 \ab)^{-1} \propto B/n$.  At large $\alpha$, a clean, low-temperature electron gas becomes unstable with respect to the formation of a CDW or Wigner crystal.  In our experiments $\alpha$ is as large as $12$, which is two orders of magnitude larger than in previous high-field experiments in STO.\cite{allen_conduction-band_2013, lin_fermi_2013} At such large values of $\alpha$, Hartree-Fock calculations predict a CDW gap that exceeds the Fermi energy, indicating a strong instability toward a spatially inhomogeneous phase.\cite{fukuyama_cdw_1978, gerhardts_magnetic_1980}

\begin{figure*}[htb]
\centering
\includegraphics[width=0.75 \textwidth]{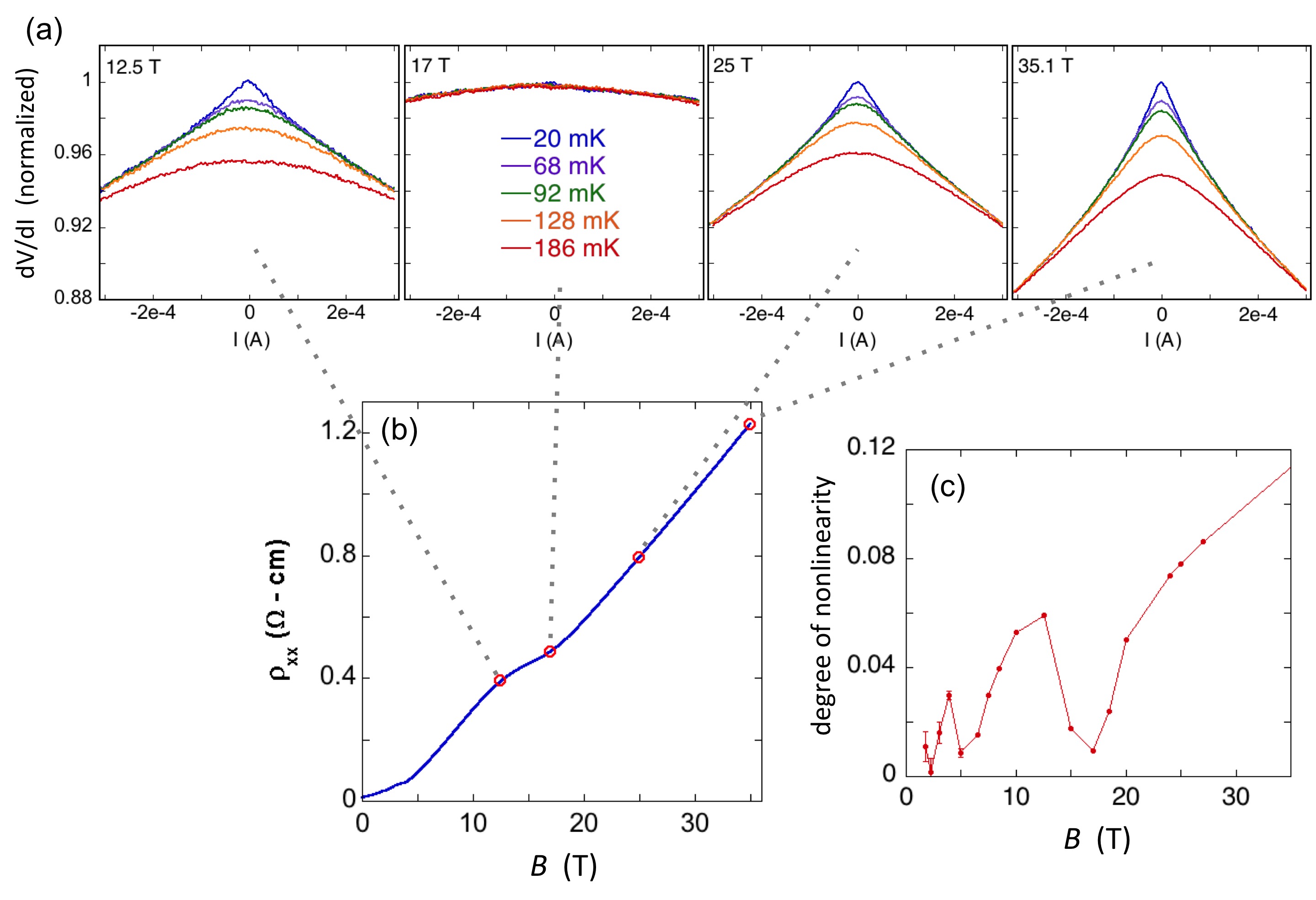}
\caption{Nonlinearity of the resistance in the EQL.  (a) The differential resistance, $dV/dI$, is plotted as a function of the source-drain bias current $I$ for different temperatures (different curves on each plot) and for different values of the magnetic field (different plots, each labeled by the corresponding value of $B$).  For each plot, the values of $dV/dI$ are normalized relative to the value at zero bias and $T = 20$\,mK.  The dotted lines indicate the value of the field relative to the phase of the SdH oscillations, which are shown in (b). (c) The degree of nonlinearity in the $I$-$V$ characteristics is shown to oscillate as a function of $B$, such that the transport is most nonlinear at maxima of the SdH oscillations, while at minima of the SdH oscillations the nonlinearity essentially disappears.  The $y$ axis corresponds to the quantity $(dV/dI|_{I=0} - dV/dI|_{I=30\,\mu\textrm{A}})/(dV/dI|_{I=0})$ at $T = 20$\,mK, which approximates the slope of the top curves in Fig.\ \ref{fig:nonlinearity}(a).}
\label{fig:nonlinearity}
\end{figure*}

Deep within the EQL, we observe a significant nonlinearity in the current-voltage ($I$-$V$) characteristics, as illustrated in Fig.\ \ref{fig:nonlinearity}(a).  Such nonlinearity typically implies ``pinning" or trapping of carriers by a disorder potential.  The observed nonlinearity is also weaker at higher temperatures, while the zero bias resistance is lower, implying weaker pinning as temperature is increased.  A careful examination of the power dissipation at low bias confirms that the nonlinearity is not a result of Joule heating, particularly at low bias (see SI Sec.\ I).  We also find it unlikely that the nonlinearity arises from scattering by domain walls separating tetragonal domains, which form below $T = 105$\,K as STO undergoes a transition from cubic to tetragonal crystal symmetry.\cite{cowley_lattice_1964}  The nonlinearity that we measure is clearly dependent on the magnetic field strength, while such a domain structure is $B$-independent.  We also observe no sign of any anomalies in the resistivity at $T = 105$\,K (see SI Sec.\ V).  

Strikingly, the observed nonlinearity appears in a re-entrant way at lower magnetic fields, becoming most pronounced at relative resistivity maxima and disappearing at relative resistivity minima.  This is illustrated in Fig.\ \ref{fig:nonlinearity}(c).  It is worth noting that a similar re-entrant nonlinearity has been observed in both NbSe$_3$ and InAs, although in both cases observations were limited to relatively high Landau levels.  In NbSe$_3$ the results were interpreted in terms of CDW physics,\cite{richard_nonlinear_1987} while in InAs it was explained in terms of changes in the effective electron temperature.\cite{bauer_low-temperature_1972}  A closer analysis of the nonlinearity is presented in SI Sec.\ V, including the scaling of the resistivity with bias voltage and magnetic field.  

Strong nonlinearity in the transport is expected when electrons form a spatially-correlated state, such as a CDW or Wigner crystal, that can be easily pinned by a disorder potential.\cite{gruner_nonlinear_1983} Such a spatially inhomogeneous state would also be consistent with the observed anisotropy in conductivity, and, indeed, previous studies of Hg$_{1-x}$Cd$_{x}$Te have interpreted similar nonlinearity as evidence for a Wigner crystal.\cite{field_evidence_1986} Working against this interpretation, however, is the very small absolute magnitude of the Coulomb interaction strength between individual electrons, owing to the large dielectric constant.  Indeed, theoretical estimates of the critical temperature $T_c$ for CDW formation\cite{fukuyama_cdw_1978, gerhardts_magnetic_1980, heinonen_electron-phonon_1986} suggest that $T_c \approx 5$\,mK or lower in our samples, and this is below our lowest measurement temperature of $20$\,mK.  Thus, if the nonlinearity in our measurements indeed arises from a CDW or Wigner crystal phase, then it likely must be understood in combination with structural distortions in the STO lattice\cite{johannes_fermi_2008, cowley_phase_1996} rather than as a simple electronic instability.

\begin{figure}[htb]
\centering
\includegraphics[width=0.48 \textwidth]{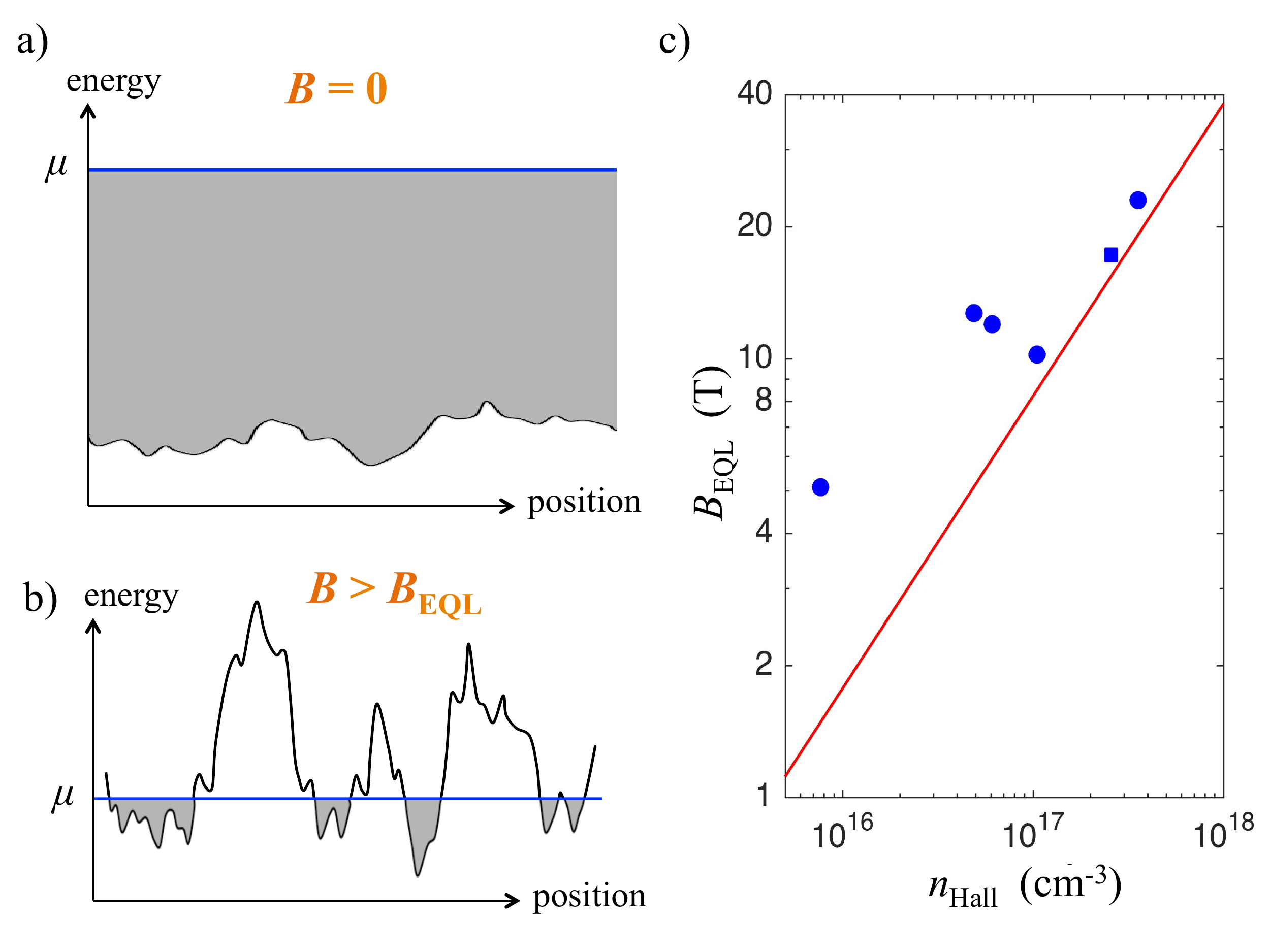}
\caption{Proposed breakup of the electron liquid into puddles at large magnetic field.  (a) At zero magnetic field, the disorder is weak compared to the typical Fermi energy, and so the density is nearly uniform spatially.  Here, the shaded area represents filled energy levels and the blue line indicates the position of the Fermi level $\mu$.  (b) Deep in the EQL, the Fermi energy is greatly reduced, and electrons reside in disconnected puddles that are localized in minima of the disorder potential.  (c) Evidence for puddling can be seen in the relatively large value of the quantum-limiting field $\beql$ for samples with low values of the electron density $n$.  Symbols represent measured values of $\beql$ (with the square symbol indicating a measurement in the $111$ direction, as in Fig.\ \ref{fig:EQL}(d)).  Experimental uncertainty in $\beql$ is $\approx 0.2$\,T, smaller than the symbol sizes.  The solid line shows the predicted value of $\beql$ for a uniform electron gas, Eq.\ (\ref{eq:BEQL}).}
\label{fig:puddling}
\end{figure}

The picture of CDW-type order also ignores the role of charged impurities, which can be expected to overwhelm electron-electron interaction effects when the carrier concentration is low.  An alternative picture, then, is to assume that the state of the electron system at large $B$ is dictated by fluctuations in the disorder potential.  In particular, if one assumes that the impurities are arranged in a spatially uncorrelated way, then random fluctuations in their local density give rise to a disorder potential with relatively large magnitude\cite{shklovskii_localization_1973, shklovskii_electronic_1984}  
\be 
\gamma = e^2 N_i^{2/3}/(4\pi\varepsilon_0\e n^{1/3}).
\label{eq:gamma}
\ee
In our samples, $\gamma$ is in the range $10$\,$\mu$eV to $20$\,$\mu$eV.  At $B = 0$, this disorder represents a relatively small perturbation to the electron Fermi level, as illustrated in Fig.\ \ref{fig:puddling}(a).  Deep within the EQL, however, the Fermi energy falls dramatically, and one can expect that the electron liquid breaks up into disconnected puddles that are localized in wells of the disorder potential [see Fig.\ \ref{fig:puddling}(b)].

In principle, this puddled state at large $B$ corresponds to an insulator, with a finite activation energy $E_a$ for electron transport that is related to the amplitude of the disorder potential.
However, owing to the large dielectric response and the relative ease of percolation between wells of the potential in three dimensions, for our samples the expected value of $E_a$ is relatively small, $E_a \approx 0.15 \gamma \approx k_B \times (15$\,mK),\cite{shklovskii_electronic_1984, aronzon_magnetic-field-induced_1990, skinner_why_2012} below our lowest measurement temperatures.  In this sense the picture of electron puddles is not inconsistent with the absence of a sharp upturn in the resistivity at large fields.  Clear observation of a magnetic field-driven metal-insulator transition in STO may require lower temperatures or samples with lower electron concentration.  It is also possible that impurity positions have some degree of spatial correlation,\cite{merkle_defect_2003, szot_localized_2002\ignore{, cuong_oxygen_2007,siegel_structure_1979}, schie_simulation_2014} which would further reduce $E_a$ by diminishing the amplitude of disorder potential fluctuations.

Support for the picture of electron puddling at large field can be found by examining the value of the field $\beql$ at which the system enters the EQL.  For a spinless, spatially uniform electron gas, 
\be 
\beql = \left(2 \pi\right)^{1/3} \Phi_0 n^{2/3},
\label{eq:BEQL}
\ee 
and one can therefore expect $\beql$ to decrease with electron density as $\beql \propto n^{2/3}$.  We find, however, that $\beql$ far exceeds this theoretical value for our low-density samples with $n \lesssim 10^{17}$\,cm$^{-3}$, and appears to saturate at a value on the order of $\approx 10$\,T.  Indeed, for our lowest-density samples, the disagreement between the observed value of $\beql$ and the prediction of Eq.\ (\ref{eq:BEQL}) is larger than $3$ times, as shown in Fig.\ \ref{fig:puddling}(c).  It should be emphasized that this disagreement is much larger than the inaccuracy that results from taking the large $N$ limit in Eq.\ (\ref{eq:SdH}), which is only $\approx 30\%$ when applied to $N = 1$.

This discrepancy between the measured and predicted values of $\beql$ can be explained within the picture of electron puddles, since the typical concentration $n_p$ of electrons within puddles in the EQL is markedly different from the volume-averaged electron concentration $n$.  Indeed, in the EQL the typical concentration of electrons within puddles takes a value that is independent of the volume-averaged electron concentration\cite{shklovskii_localization_1973, aronzon_magnetic-field-induced_1990}:
\be 
n_p = \left( \frac{\pi^4}{2} \right)^{3/14}  \frac{(\ab/\lb)^{2/7}}{(N_i \lb^2 \ab)^{5/7}} N_i.
\label{eq:np}
\ee 
The value of $n_p$ is determined by statistical fluctuations of the disorder potential over length scales much shorter than the correlation length of the potential.  Such fluctuations are driven by the random influence of the many impurity charges, rather than by the weak nonlinear screening from the sparse electrons, and this leads to the independence of $n_p$ on $n$ implied by Eq.\ (\ref{eq:np}).\cite{shklovskii_localization_1973, aronzon_magnetic-field-induced_1990}

For our experiments, Eq.\ (\ref{eq:np}) suggests that $n_p$ is on the order of a few times $10^{17}$\,cm$^{-3}$ in the EQL, while the typical puddle radius is $\sim 40$\,nm.  One can then arrive at an estimate for $\beql$ by inserting $n_p$ from Eq.\ (\ref{eq:np}) into Eq.\ (\ref{eq:BEQL}).  This procedure gives $\beql \approx 10 (\Phi_0/\ab^2) (N_i \ab^3)^{4/9}$, which for our samples is on the order of $10$\,T.  This is consistent with the observed value in samples with small $n \lesssim 10^{17}$\,cm$^{-3}$.  At larger values of the doping, the carrier concentration $n$ presumably becomes comparable to the total impurity concentration $N_i$, and the electron gas becomes uniform again so that Eq.\ (\ref{eq:BEQL}) is valid.

One can also rationalize the nonlinearity of the $I$-$V$ characteristics at $B > \beql$ within the picture of electron puddles.  In particular, an applied electric field facilitates the thermal activation of electrons between adjacent puddles by adding a contribution to the potential energy that varies linearly with the electron position. This picture implies a characteristic electric field scale that is consistent with the observed nonlinearity in the EQL (see SI Sec.\ V).  The nonlinearity is also identical in both the parallel and perpendicular field directions, which is again consistent with the puddle picture (SI Sec.\ V).  Still, the oscillatory behavior of the nonlinearity presented in Fig.\ \ref{fig:nonlinearity} remains a prominent puzzle.  

Finally, the picture of electron puddles is also qualitatively consistent with the observed linear MR in the EQL, since it corresponds to a landscape where both the disorder potential and the electron concentration have strong spatial fluctuations with a long correlation length.  The observed anisotropy of resistivity is also typical in such situations.\cite{shayegan_magnetic-field-induced_1985, aronzon_magnetic-field-induced_1990} (It should be noted, however, that many scenarios for electron transport in the EQL produce significant anisotropy,\cite{askerov_electron_1994} and therefore the anisotropy by itself cannot generally be used as a strong diagnostic of the electron state.)


In conclusion, this paper has presented a clear experimental demonstration of the EQL in bulk STO across a range of samples.  Our experiments probe much more deeply into the EQL than previous studies of STO, and into the regime of small $n \lb^2 \ab$, for which a charge ordering instability has long been predicted to occur for a three-dimensional electron gas.  While some of our measurements are consistent with the formation of such a CDW state, including in particular a striking nonlinearity in the $I$-$V$ characteristics, it seems unlikely to us that electron-electron interactions alone are sufficient to induce such a state within the regime of our experiments.  Nonetheless, it remains an open question whether a field-induced CDW state can result from the combination of electron interactions and structural distortions in the STO lattice.  An alternate explanation for our measurements is that the large-field behavior is dominated by puddling of electrons in minima of the disorder potential.  The observed saturation of the quantum limiting field is apparently consistent with this picture.  It may be worth considering, however, whether CDW physics can coexist with an electron-puddled structure.  Future studies may provide important additional insight into the electron state in the EQL, particularly if they can complement our transport measurements with thermodynamic probes like capacitance or tunneling spectroscopy, or with studies of magnetic field-driven structural transitions.

\begin{small}
\vspace*{2ex} \par \noindent
{\em Acknowledgments.}

Initial measurements of quantum oscillations in reduced STO samples were carried out at the NHMFL in Los Alamos by AB with J.\ Singleton and F.\ Balakirev.  We are grateful to C.\ Leighton, P.\ B.\ Littlewood, A.\ Lopez-Bezanilla, B.\ I.\ Shklovskii, and K.\ V.\ Reich for helpful discussions.  AB  acknowledges the support of the U.S. Department of Energy (DOE), Office of Science, Basic Energy Sciences (BES), Materials Sciences and Engineering Division.  The use of facilities at the Center for Nanoscale Materials, was supported by the U.S. DOE, BES under contract No.\ DE-AC02-06CH11357.  Theory work by BS was initially supported at Argonne National Lab by the U.S. Department of Energy, Office of Science, under contract no. DE-AC02-06CH11357; subsequent theory work was supported as part of the MIT Center for Excitonics, an Energy Frontier Research Center funded by the U.S. Department of Energy, Office of Science, Basic Energy Sciences under Award no. DE-SC0001088.   
The NHMFL is supported by the NSF Cooperative Agreement No.\ DMR-1157490 and the State of Florida.

\end{small}

%

\widetext
\clearpage
\begin{center}
\textbf{\large Supplementary Information for ``Spatially inhomogeneous electron state deep in the extreme quantum limit of strontium titanate''}
\end{center}

\date{\today}

\renewcommand{\theequation}{S\arabic{equation}}
\renewcommand{\thefigure}{S\arabic{figure}}
\renewcommand{\bibnumfmt}[1]{[S#1]}
\renewcommand{\citenumfont}[1]{S#1}
\setcounter{figure}{0}

\section{Sample preparation and measurement setup}
\label{sec:sample}

\emph{Sample Preparation:}\;
As mentioned in the main text, the SrTiO$_3$ single crystals studied here were obtained from Crys Tec GmbH.\cite{crystecSI} On one side the samples were  atomically smooth with regular unit cell high terraces, while the opposite face was unpolished. The (002) peak in all samples in this study had rocking curve full width at half maxima in the range $0.027^\circ < \Delta \omega < 0.041^\circ$. The samples were first cleaned with solvents (including trichloroethylene) to remove any residue or particulate and were then mounted on a stainless steel sample holder. All annealing was carried out using a SiC radiative heater. The samples were annealed in a prep chamber at $400\, ^\circ$C for approximately 30 min to get rid of any adsorbed water or solvents on the holder and substrate.  After the pressure of the prep chamber dropped below $4 \times 10^{-6}$\,Pa ($3 \times 10^{-8}$\,Torr), the sample was cooled down and inserted into the main vacuum chamber, which was typically at a base pressure between $7 \times 10^{-8}$\,Pa and $3 \times 10^{-7}$\,Pa ($5 \times 10^{-10}$\,Torr and $2 \times 10^{-9}$\,Torr). The sample was heated to temperatures between $680\,^\circ$C and $850\,^\circ$C, depending on the doping level desired, at a rate of $30\,^\circ$C/min to $40\,^\circ$C/min. The pressure in the chamber near the end of the annealing process was $\approx 3 \times 10^{-6}$\,Pa ($2 \times 10^{-8}$\,Torr), comprising mostly H$_2$. Samples were held at the annealing temperature for exactly $60$ min, after which time the power to the heater was rapidly reduced and then turned off and the sample was allowed to cool, which took approximately $30$ min. 

\

\emph{Contacts:}\;
The current contacts spanned the full width of the sample ($\approx 0.5$\,mm $\times$ 7.5\,mm), while four voltage contact tabs (with dimensions $~ 1.5$\,mm $\times$ $1.5$\,mm) were deposited at the edges of the sample (spaced $\approx 1.8$\,mm apart). NiCr ($50$\,nm)/Au ($200$\,nm) contacts were sputtered on the unpolished side of the sample after Ar ion milling only the contact pad area for $\approx 15$\,min ($400$\,V, $30$\,mA, incident at $\approx 45^\circ$). All samples were typically measured within a week of annealing, as the resistivity was found to increase progressively with time. 

\

\begin{figure}[htb]
\centering
\includegraphics[width=0.5 \textwidth]{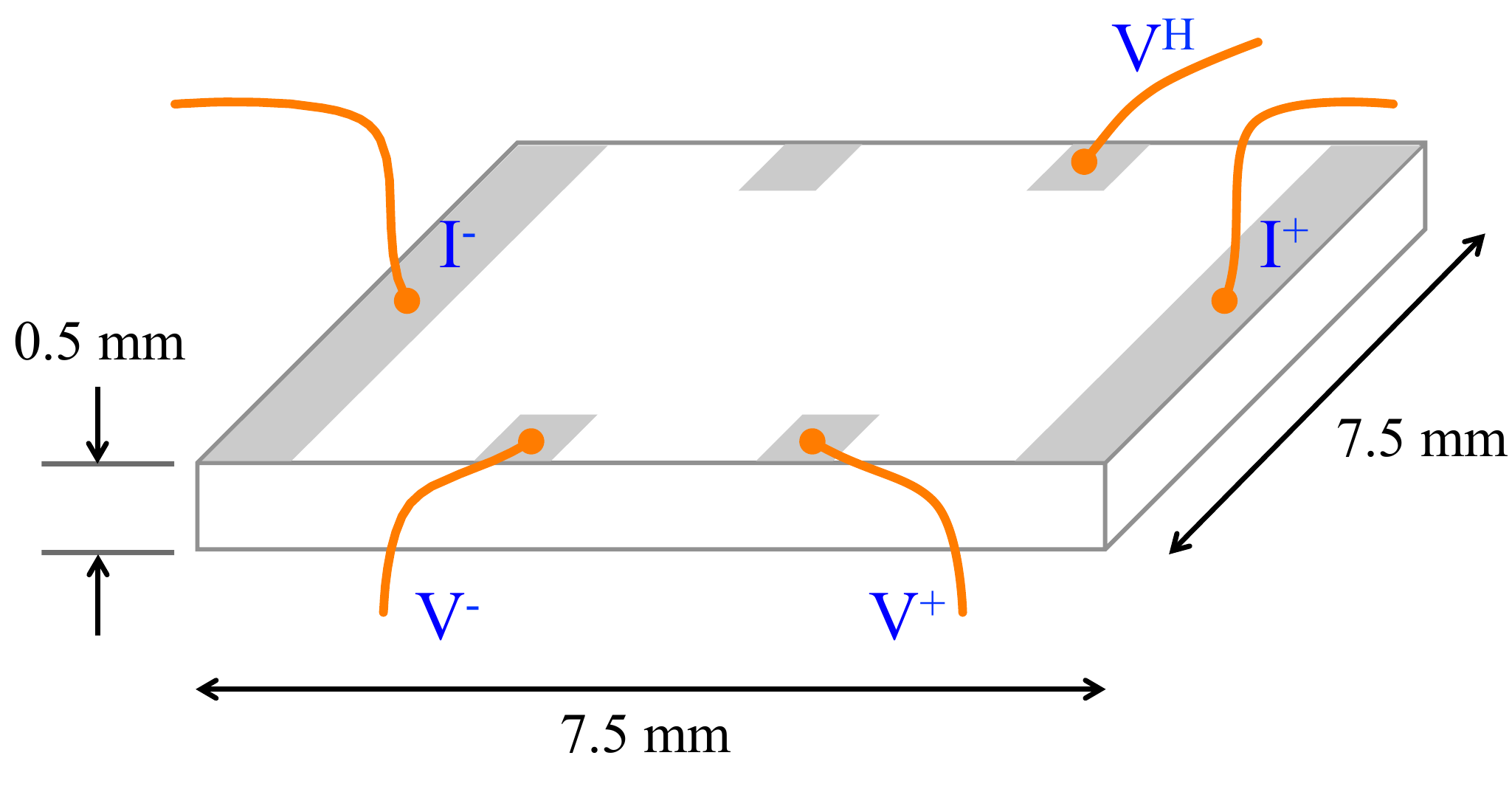}
\caption{Schematic of the measurement setup, together with physical dimensions of the sample.}
\label{fig:sample}
\end{figure}

\emph{Transport Measurements:}\;
Simultaneous measurements of the longitudinal and Hall voltages, $V_{xx}$ and $V_{xy}$, respectively, were carried out in a 5-terminal Hall geometry, as illustrated in Fig.\ \ref{fig:sample}.  Our measurements used the direct current reversal technique, and the experimental setup was an improved version of the system described in Ref.\ \onlinecite{suslov_stand_2010}.  

Utilized current sources and nanovoltmeters 
have a built-in option for measurements of differential conductance or differential resistance.  Thus, we used the same setup for measurements of differential resistance $dV_{xx}/dI$ in the 4-terminal geometry.  (Elsewhere in this article, the notation $dV/dI$ is used in place of $dV_{xx}/dI$.)  The differential resistance at a given bias current, $\ibias$, was characterized by measuring changes in voltage, $dV$, associated with small excursions in the current, $dI$, about the applied bias.  During the measurements $\ibias$ was swept through a specified range with a finite step size $\Delta \ibias$.  

Experiments were performed at several magnet/cryostat configurations available at the National High Magnetic Field Laboratory, and the study was performed in magnetic field as high as $45$\,T and at temperatures ranging from $20$\,mK to $200$\,mK using dilution cryostats.  Care was taken to ensure that our measurements were not affected by Joule heating.

Contact resistance at low temperatures was too low to be measured reliably ($1\,\Omega$ or lower). Fabrication of contacts with such low resistance was critical for carrying out low-noise voltage measurements at low temperatures, where the samples typically had resistance values in the range $0.1$\,$\Omega$ to $100$\,$\Omega$. Furthermore, for measurements in the dilution fridge, the low contact resistances at the current contacts were essential for minimizing heating due to the excitation current.

\

\begin{figure*}[htb]
\centering
\includegraphics[width=0.8 \textwidth]{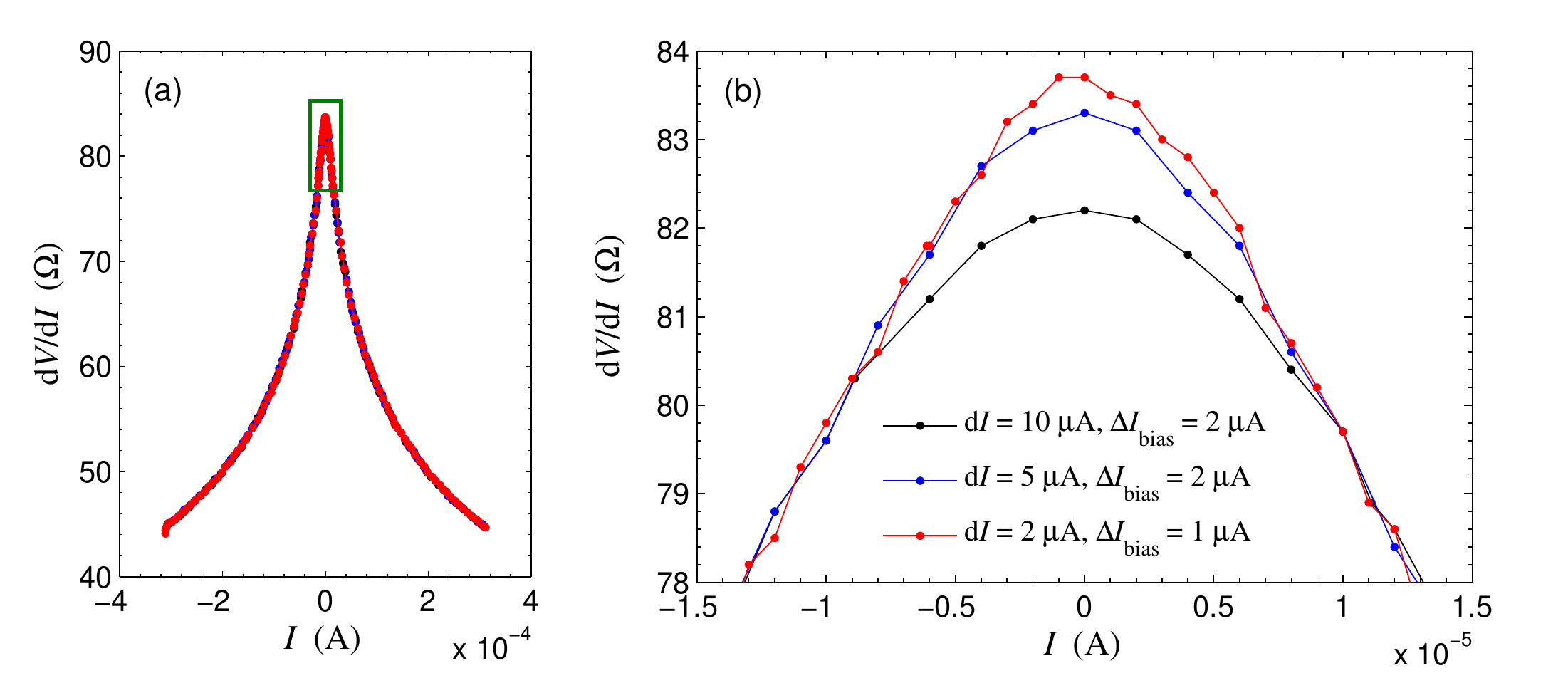}
\caption{Differential resistance measured at temperature $20$\,mK and magnetic field $45$\,T, shown for different values of the excitation current $dI$ and the current step size $\Delta \ibias$.  (a) Shows the measured differential resistance over a wide range of bias current.  (b) Shows the same data plotted very near the point of zero bias.  [The rectangle in (a) indicates the approximate range and domain of (b).]}
\label{fig:heating}
\end{figure*}

\emph{Heating:}\;
In order to characterize the effects of the excitation current on heating in the sample, and on the value of the measured resistance in our experiments, we measured $dV/dI$ for a range of currents between $-300\,\mu$A and $+300\,\mu$A for several values of the current step size $\Delta \ibias$ and the excitation current $dI$. By varying these values at low $\ibias$ we were able to substantially vary the power dissipated in a $dV/dI$ measurement. We observed that for $\ibias > 10\,\mu$A, the value of $dV/dI$ was independent of the utilized measurement parameters.  This is shown for $dI = 2\,\mu$A, $5\,\mu$A, and $10\,\mu$A and $\Delta  \ibias = 1\,\mu$A, $2\,\mu$A and 2$\,\mu$A, respectively, in Fig.\ \ref{fig:heating}. Our estimations show that at $\ibias = 10\,\mu$A, current step size $\Delta  \ibias = 1\,\mu$A, and excitation current $dI = 2\,\mu$A the power dissipated in the sample was about $57$\,nW, whereas for $\ibias = 10\,\mu$A, $\Delta \ibias = 2\,\mu$A, and $dI = 5\,\mu$A this power was about $130$\,nW. Thus, while corresponding values of the dissipated power differ by more than a factor of $2$, the measurements nonetheless produce the same result for the differential resistance. Therefore, at least at low values of the current bias, the change in $dV/dI$ measured as a function of bias (Fig.\ 4 of the main text) is not due to heating. The disagreement at zero bias between the curves for $dI = 2\,\mu$A, $5\,\mu$A, and $10\,\mu$A in Fig.\ \ref{fig:heating} can be explained by a coarse graining effect produced by the larger $dI$ values near the sharp cusp-like feature at zero bias.  The data for $dI = 2\,\mu$A is closest to the ``intrinsic'' curve.

\section{Zero-field mobility and estimate of impurity concentration}

\label{sec:mobility}

As mentioned in the main text, as-grown STO crystals generally have a relatively large number of impurities that act as deep acceptors.\cite{de_souza_behavior_2012SI, rice_persistent_2014}   We therefore expect that the concentration of impurities $N_i$ significantly exceeds the concentration $n$ of free electrons.  Here we show that this expectation is consistent with measurements of the zero-field, low-temperature mobility.

In particular, at low enough temperatures that the mobility saturates at a constant value, one can expect that the electron mobility $\mu_e$ is limited primarily by scattering from ionized impurities (Rayleigh scattering).  For such scattering processes, screening of the impurity potential by conduction electrons is essentially irrelevant in our samples.  This can be seen by examining the Thomas-Fermi screening radius
\be 
r_s = \sqrt{\frac{\e_0 \e}{e^2 \nu}},
\ee 
where 
$\nu$ is the electron density of states. In our samples $r_s \propto \sqrt{a_B/n^{1/3}} \sim 60$\,nm, which is much longer than the distance between electrons or between charged impurities. (Here, $\ab$ denotes the effective Bohr radius at zero magnetic field.)  Consequently, the screening of impurities can be ignored for calculating the scattering cross section.
In this case,\cite{dingle_scattering_1955}
\be 
\mu_e = \frac{9 \pi e n \ab^2}{2 \hbar N_i \ln(3 \pi^5 n \ab^3)}.
\label{eq:Rayleigh}
\ee 

Solving Eq.\ (\ref{eq:Rayleigh}) for $N_i$ gives an estimate of the impurity concentration for a given mobility $\mu_e$ and carrier density $n$.  For the samples presented here, this estimate yields values of $N_i$ of a few times $10^{18}$\,cm$^{-3}$, which is consistent with our assumptions in the main text and with previous studies.\cite{spinelli_electronic_2010, rice_persistent_2014, de_souza_behavior_2012SI}

Of course, this value can be considered an upper-bound estimate for $N_i$, since the presence of other, short-ranged scatterers will also decrease the mobility.  The Coulomb potential created by impurities is also not perfectly described by a constant dielectric function $\e$, since the dielectric function is dispersive and thus the dielectric response is not fully developed at short distances from the impurity.  Hence, scattering by impurities may be somewhat stronger than implied by Eq.\ (\ref{eq:Rayleigh}), which would further reduce the estimate for $N_i$.  Theoretical estimates for the dispersive nature of the dielectric function, however, suggest that this effect is not too large for isolated monovalent charges.\cite{reich_accumulation_2015}

\

\section{Hall Resistance}
\label{sec:Hall}

\begin{figure}[htb]
\centering
\includegraphics[width=0.4 \textwidth]{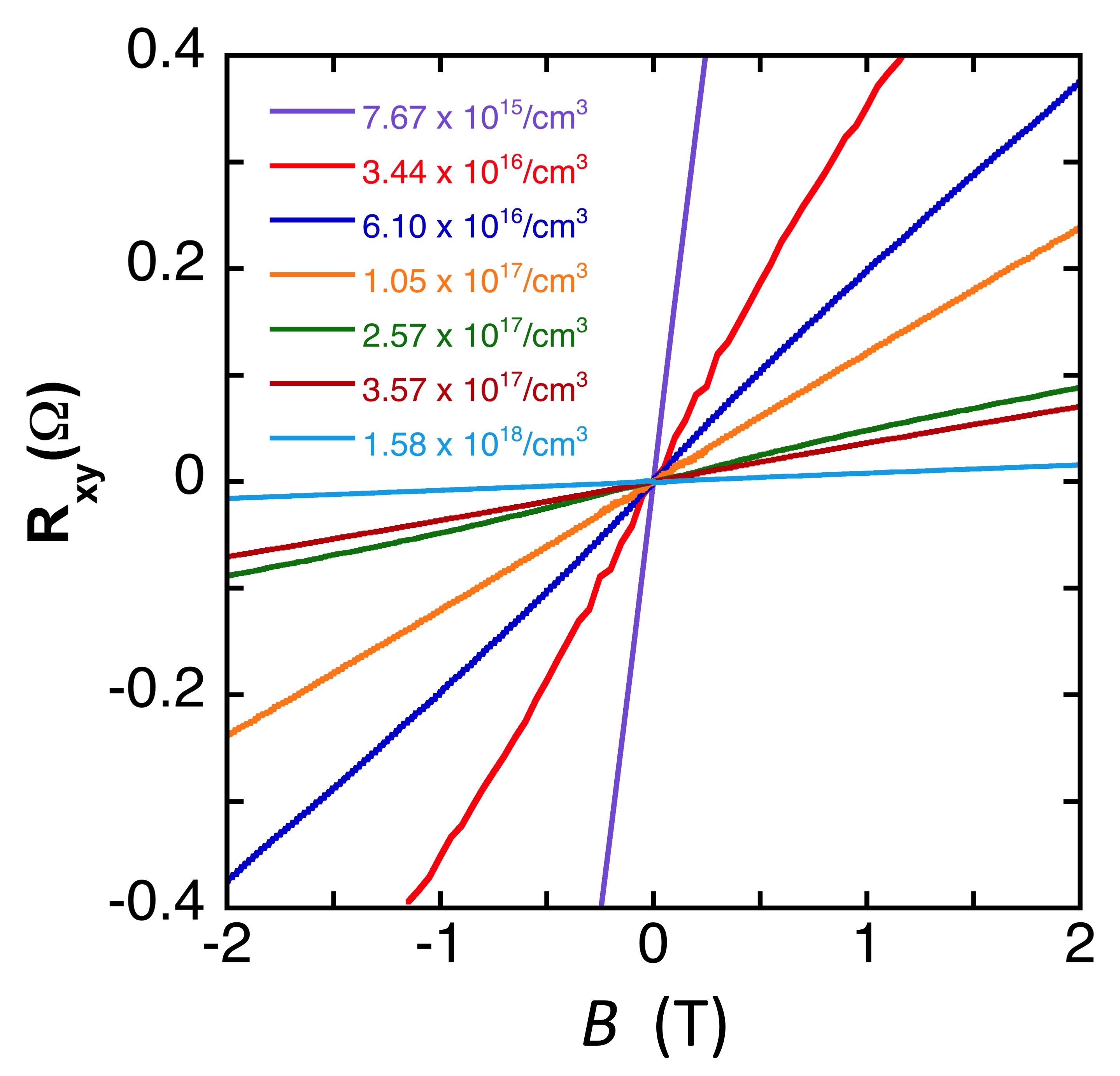}
\caption{Low temperature Hall data for the series of doped SrTiO$_3$ samples studied in this work.  All samples are $n$-type, and curves are labeled by the extracted value of the Hall carrier density.  All measurements were taken at $T = 25$\,mK, with the exception of those corresponding to $n_\text{Hall} = 3.57 \times 10^{17}$\,cm$^{-3}$ and $n_\text{Hall} = 7.67 \times 10^{15}$\,cm$^{-3}$.  The former was measured at $T = 300$\,mK, and the latter was measured at $T = 2$\,K.}
\label{fig:Hall}
\end{figure}

In Fig.\ \ref{fig:Hall} we present Hall resistance measurements for the samples studied in this work. 

\

\section{Shubnikov-de Haas Analysis}
\label{sec:SdH}

In the main text are presented the results of a Shubnikov-de Haas (SdH) analysis of our doped STO samples.  In particular, in this analysis we identify oscillations of the longitudinal resistivity $\rho_{xx}$ that are periodic in $1/B$.  As mentioned in the main text, this is done by subtracting a smooth, fourth-order polynomial from the curve $\rho_{xx}(B)$ and identifying the field values $B_N$ corresponding to the maxima of the background-subtracted curve $\Delta \rho_{xx}(B)$.  

\begin{figure*}[htb]
\centering
\includegraphics[width = 0.95 \textwidth]{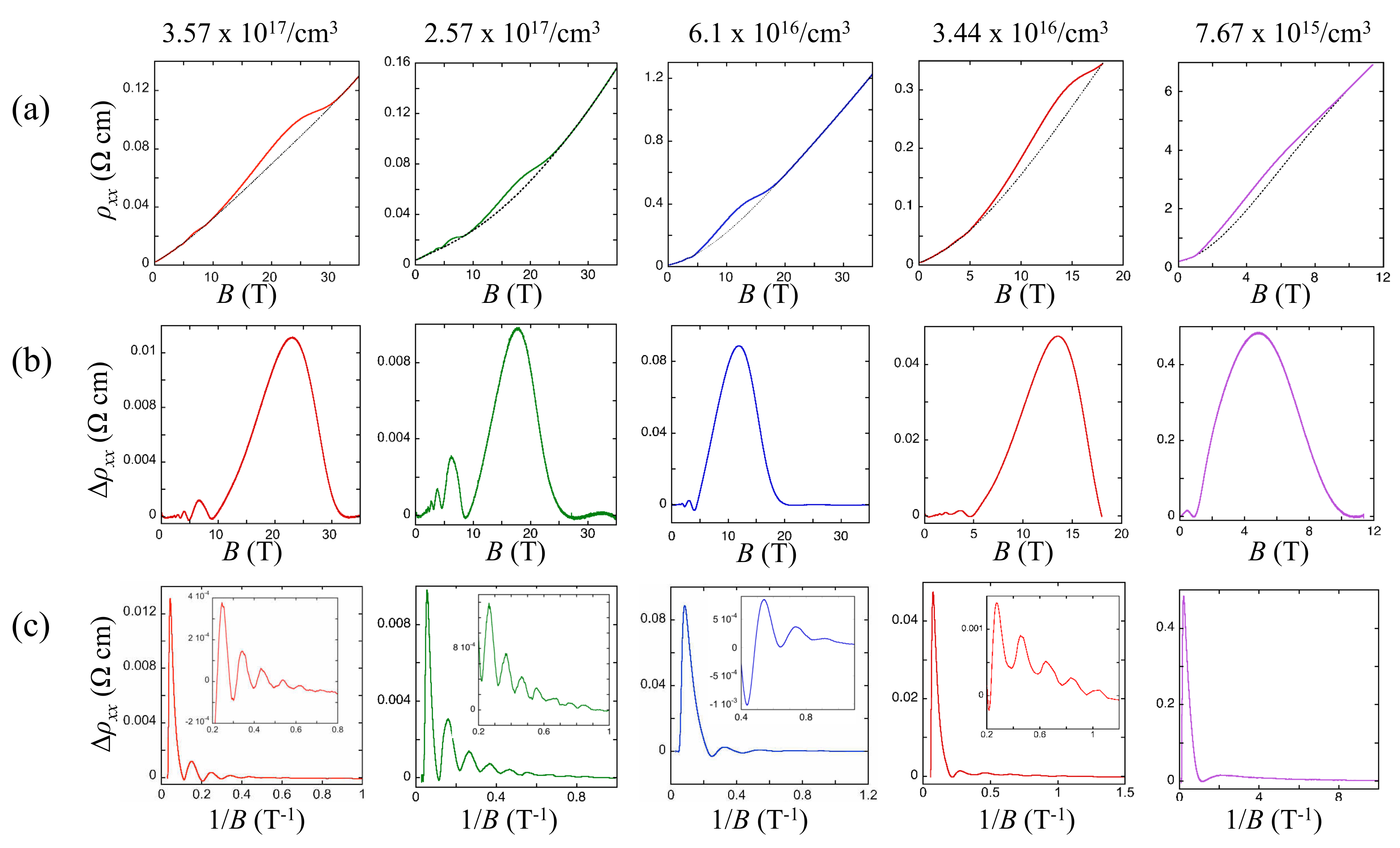}
\caption{Longitudinal resistivity ($\rho_{xx}$) vs.\ magnetic field ($B$) at low temperatures for five of the samples studied in this work. For each sample, $\rho_{xx}$ vs $B$ is plotted in (a), where the dotted line depicts a fourth order polynomial which serves as a smooth background. The background subtracted data $\Delta \rho_{xx}$ are plotted vs.\ $B$ in (b) and vs.\ $1/B$ in (c), where quantum oscillations of the resistivity are clearly visible.  Insets in (c) show the same data plotted with a smaller range of the vertical axis, so that lower-amplitude oscillations are visible.  Each column of plots is labeled by the corresponding value of the Hall density $n_\text{Hall}$; columns are arranged in order of decreasing $n_\text{Hall}$.  All measurements were taken at $T = 25$\,mK, with the exception of the sample having $n_\text{Hall} = 3.57 \times 10^{17}$\,cm$^{-3}$, which was measured at $T = 300$\,mK.
}
\label{fig:SdH}
\end{figure*}

In Fig.\ \ref{fig:SdH} we show full details of this data analysis procedure for five of our studied samples, each of which is driven into the extreme quantum limit at large field.  In particular, for each sample we show the raw curve $\rho_{xx}(B)$, the background-subtracted curve $\Delta \rho_{xx}(B)$, and the same data $\Delta \rho_{xx}$ plotted against the inverse field $1/B$.  

\

\section{Nonlinear transport}
\label{sec:pinningfield}

\begin{figure*}[htb]
\centering
\includegraphics[width=0.7 \textwidth]{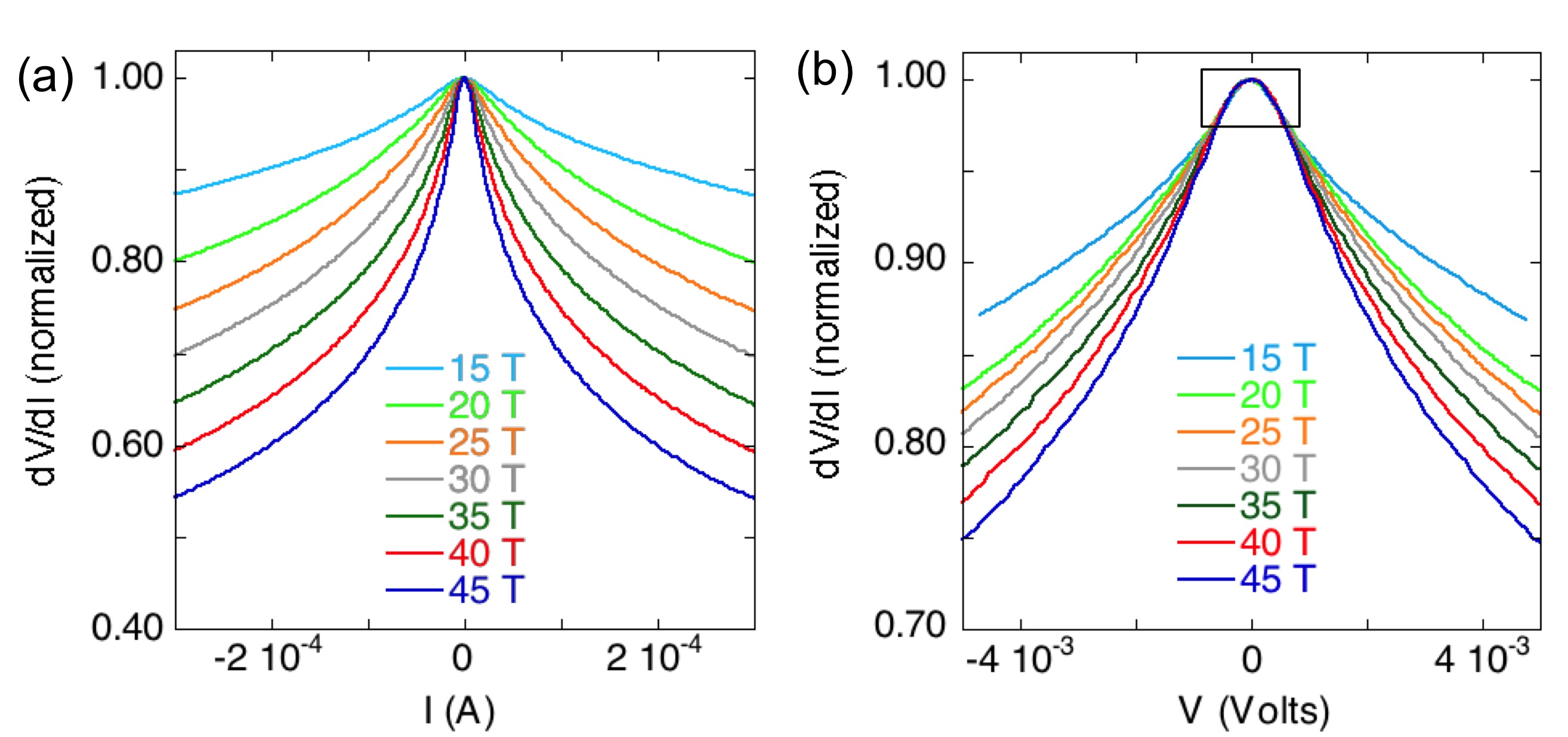}
\caption{Plots of the normalized differential resistance, $(dV/dI) / (dV/dI)|_{I = 0}$, versus (a) bias current $I$ and (b) bias voltage $V$ at $T = 20$\,mK for a range of magnetic field values.  The nonlinearity becomes stronger as the magnetic field is increased.  For small values of the bias voltage $V$, the curve $dV/dI$ versus $V$ acquires a nearly universal shape, independent of the magnetic field strength.}
\label{fig:dIdVscaling}
\end{figure*}

\emph{Resistivity Scaling:}\;
At high magnetic fields, the resistivity $\rho$ exhibits a significant nonlinearity, such that $\rho$ is a function of both the bias voltage $V$ and the magnetic field $B$.  We observe that the resistivity can in fact be scaled in such a way that different curves for the differential resistance, $dV/dI$, as a function of $V$ collapse on top of each other at small $V$.  This is shown in Fig.\ \ref{fig:dIdVscaling}(b).  The collapse of the curves suggests that one can write $dV/dI = f(V)h(B)$, where $f$ and $h$ are scaling functions.   

From the data in Fig.\ \ref{fig:dIdVscaling} one can extract a characteristic electric field scale $F_0$ for the nonlinearity in multiple different ways.  For example, one can define $F_0$ as the value of the electric field above which the scaling shown in Fig.\ \ref{fig:dIdVscaling}(b) is lost.  This definition gives $F_0 \approx 10$\,mV/cm.  Alternatively, one could define $F_0$ as the value for which the differential resistance drops to half its $V = 0$ value.  Such a definition gives $F_0$ on the order of $\approx 100$\,mV/cm, depending on the value of $B$.

\

\begin{figure}[htb]
\centering
\includegraphics[width=0.5 \textwidth]{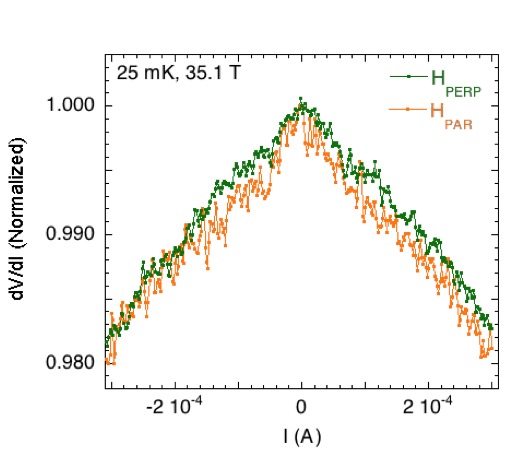}
\caption{Differential resistance, $dV/dI$, plotted as a function of the bias current $I$ and normalized to the value of $dV/dI$ at $I = 0$.  The green curve corresponds to the case where the applied magnetic field is perpendicular to the bias current, while the orange curve corresponds to the case where the magnetic field and current are parallel.  Measurements in this plot correspond to $T = 25$\,mK and $B = 35.1$\,T.}
\label{fig:parperp}
\end{figure}

\emph{Nonlinearity in the perpendicular versus parallel field directions:}\;
Our analysis of the nonlinearity has largely focused on the case where the magnetic field direction is perpendicular to the current direction.  Here we briefly show results for the case where the magnetic field and current directions are co-linear.  As one can see in Fig.\ \ref{fig:parperp}, the two cases give essentially identical results for the nonlinear differential resistance.

\

\emph{Tetragonal domain walls:}\;
At temperatures below $T = 105$\,K, STO is known to undergo a transition from cubic to tetragonal crystal symmetry.\cite{cowley_lattice_1964SI}  Consequently, at $T < 105$\,K the sample contains domain walls between differently-oriented tetragonal domains, and these can potentially influence the electron transport.  (For example, such influence has recently been studied at the STO-LaAlO$_3$ interface.\cite{kalisky_locally_2013, honig_local_2013})  These domains are typically tens of microns in size, and the domain walls have a width of about $2$\,nm and are associated with an electronic energy scale of about $3.2$\,meV.\cite{cao_landau-ginzburg_1990}

\begin{figure}[htb]
\centering
\includegraphics[width=0.5 \textwidth]{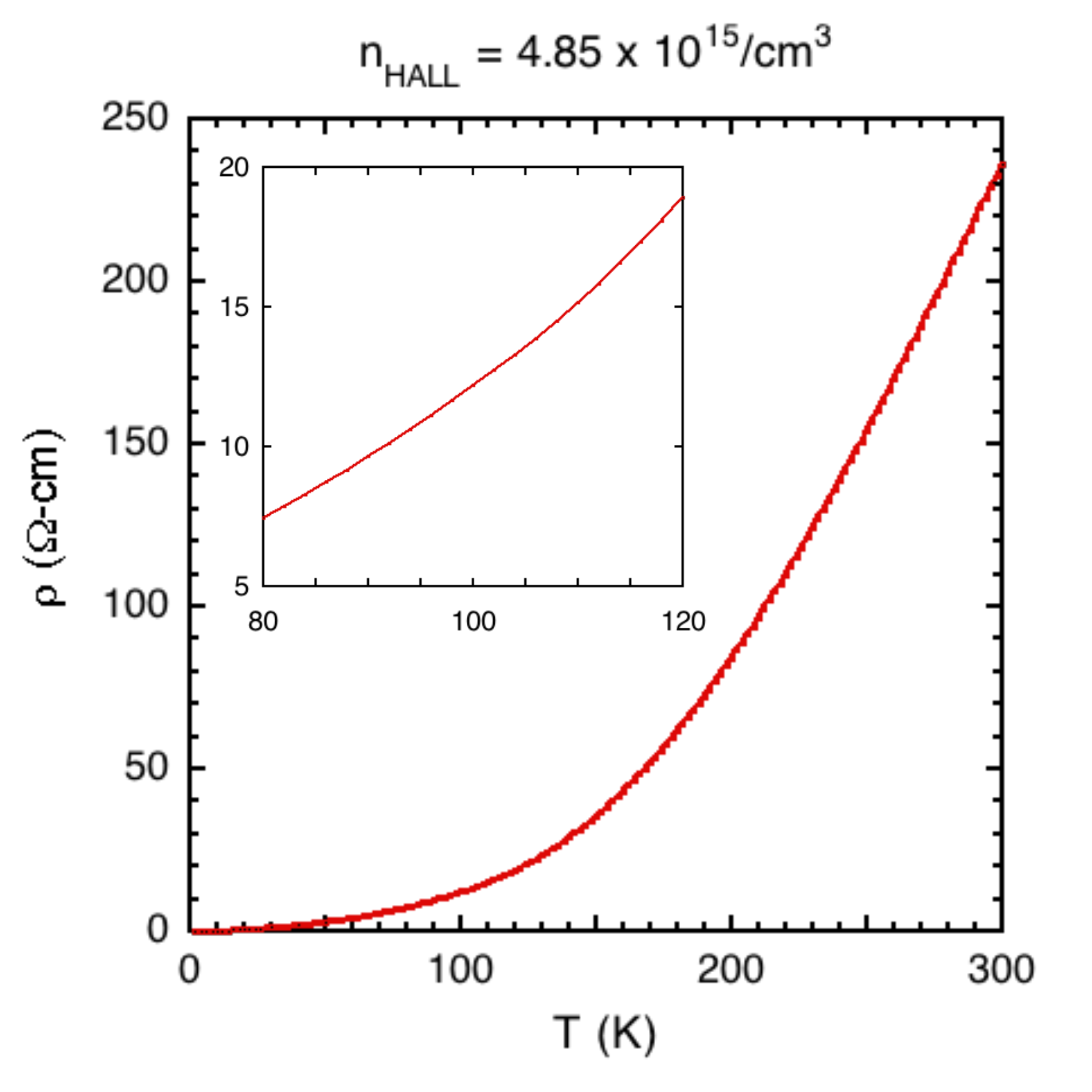}
\caption{Zero-field resistivity for our lowest-density sample as a function of temperature.  There is no sign of any anomalies in the transport at $T = 105$\,K, which corresponds to a structural transition in the STO lattice.}
\label{fig:rhoT}
\end{figure}

In our experiments, however, we find it unlikely that these domain walls are related to the observed nonlinearity in the electron transport. Most tellingly, the features we observe in the electron transport are strongly magnetic field-dependent,  while the domain structure at fixed low temperature is completely insensitive to magnetic field.  In addition, we see no sign of any anomalies in the resistivity at $T = 105$\,K, at which the structural transition occurs.  This is shown in Fig.\ \ref{fig:rhoT}.

\

\emph{Nonlinearity in the Electron Puddle Picture:}\;
In the picture of ``electron puddles'' presented in the main text, there is a natural electric field scale for the nonlinearity of the resistivity, which can be derived as follows.   

For an almost-completely-compensated semiconductor in the EQL, electrons become localized in wells of the disorder potential with typical radius\cite{ aronzon_magnetic-field-induced_1990SI}
\be 
r_p \simeq (2 \pi^4)^{2/7} (\ab \lb^6)^{1/7} (N_i \lb^2 \ab)^{1/6}
\label{eq:rp}
\ee 
and typical concentration $n_p$ given by Eq.\ (4) of the main text.
One can estimate the typical distance $R$ between puddles by noting that the total number of electrons within a puddle is equal to $Q_p \simeq (4 \pi/3) n_p r_p^3$, and therefore the volume-averaged concentration of electrons is $n \simeq Q_p/R^3$.  Rearranging this expression for $R$ gives a typical distance between puddles $R \simeq (4 \pi n_p/3n)^{1/3} r_p$.

In the absence of a bias electric field, the typical activation energy between neighboring puddles is $E_a = c \gamma$, where $\gamma$ is the typical amplitude of the disorder potential [see Eq.\ (2) of the main text] and $c$ is a numerical factor that is typically $\approx 0.15$.\cite{skinner_why_2012SI}  One can define the typical field scale $F_0$ as the value of the electric field for which the difference in electric potential between puddles due to the applied field becomes equal to the activation energy $E_a$.  In other words, $e F_0 R \simeq E_a$.  Rearranging this equation for $F_0$ and substituting $R \simeq (4 \pi n_p/3n)^{1/3} r_p$ gives
\begin{align}
F_0 & \simeq c \left(\frac{3}{4 \pi} \right)^{1/3} \frac{e N_i^{2/3}}{4 \pi \varepsilon_0 \e n_p^{1/3} r_p} 
\label{eq:F0} \\
& \simeq 0.10 c \frac{e}{4 \pi \varepsilon_0 \e} \left( \frac{N_i^3}{\ab \lb^4} \right)^{1/7}, \nonumber
\end{align}
where the second equality is reached by substituting 
Eqs.\ (4) and (\ref{eq:rp}) into Eq.\ (\ref{eq:F0}).

For our samples, Eq.\ (\ref{eq:F0}) gives $F_0 \approx 50$\,mV/cm to $80$\,mV/cm for $B$ between $15$\,T and $45$\,T.  This range can be compared to the empirical values of $F_0$, which are between $10$\,mV/cm and $100$\,mV/cm.

\end{document}